\documentclass[prl,twocolumn,letterpaper, superscriptaddress,longbibliography]{revtex4-1}
\usepackage{graphicx}
\usepackage{dcolumn}
\usepackage{bm}
\usepackage{color}
\usepackage{tabularx}
\usepackage{array}

\usepackage[utf8]{inputenc}
\usepackage[T1]{fontenc}
\usepackage{color,soul}
\usepackage{tabularx}
\usepackage{amsmath}

\makeatother
\begin{document}
\title{Phase-resolving spin-wave microscopy using infrared strobe light}

\author{Yuzan Xiong}
\affiliation{Department of Physics and Astronomy, University of North Carolina at Chapel Hill, Chapel Hill, NC 27599, USA}
\author{Andrew Christy}
\affiliation{Department of Physics and Astronomy, University of North Carolina at Chapel Hill, Chapel Hill, NC 27599, USA}
\affiliation{Department of Chemistry, University of North Carolina at Chapel Hill, Chapel Hill, NC 27599, USA}
\author{Muntasir Mahdi}
\affiliation{Department of Electrical and Computer Engineering, Auburn University, Auburn, AL 36849, USA}
\author{Rui Sun}
\affiliation{Department of Physics and Organic and Carbon Electronics Lab (ORaCEL), North Carolina State University, Raleigh, NC 27695, USA}
\author{Yi Li}
\affiliation{Materials Science Division, Argonne National Laboratory, Argonne, IL 60439, USA}
\author{Robert D. Geil}
\affiliation{Department of Chemistry, University of North Carolina at Chapel Hill, Chapel Hill, NC 27599, USA}
\author{James F. Cahoon}
\affiliation{Department of Chemistry, University of North Carolina at Chapel Hill, Chapel Hill, NC 27599, USA}
\author{Frank Tsui}
\affiliation{Department of Physics and Astronomy, University of North Carolina at Chapel Hill, Chapel Hill, NC 27599, USA}
\author{Binbin Yang}
\affiliation{Department of Electrical and Computer Engineering, North Carolina A$\&$T State University, Greensboro, NC 27411, USA}
\author{Tae Hee Kim}
\affiliation{Department of Physics, Ewha Womans University, Seodaemun, Seoul, Republic of Korea}
\author{Jia-Mian Hu}
\affiliation{Department of Materials Science and Engineering, University of Wisconsin-Madison, Madison, Wisconsin, 53706, USA}
\author{Dali Sun}
\affiliation{Department of Physics and Organic and Carbon Electronics Lab (ORaCEL), North Carolina State University, Raleigh, NC 27695, USA}
\author{Michael C. Hamilton}
\affiliation{Department of Electrical and Computer Engineering, Auburn University, Auburn, AL 36849, USA}
\author{Valentine Novosad}
\affiliation{Materials Science Division, Argonne National Laboratory, Argonne, IL 60439, USA}
\author{Wei Zhang}
\thanks{zhwei@unc.edu}
\affiliation{Department of Physics and Astronomy, University of North Carolina at Chapel Hill, Chapel Hill, NC 27599, USA}

\begin{abstract}

The needs for sensitively and reliably probing magnetization dynamics have been increasing in various contexts such as studying novel hybrid magnonic systems, in which the spin dynamics strongly and coherently couple to other excitations, including microwave photons, light photons, or phonons. Recent advances in quantum magnonics also highlight the need for employing magnon phase as quantum state variables, which is to be detected and mapped out with high precision in on-chip micro- and nano-scale magnonic devices. Here, we demonstrate a facile optical technique that can directly perform concurrent spectroscopic and imaging functionalities with spatial- and phase-resolutions, using infrared strobe light operating at 1550-nm wavelength. To showcase the methodology, we spectroscopically studied the phase-resolved spin dynamics in a bilayer of Permalloy and Y$_3$Fe$_5$O$_{12}$ (YIG), and spatially imaged the backward volume spin wave modes of YIG in the dipolar spin wave regime. Using the strobe light probe, the detected precessional phase contrast can be directly used to construct the map of the spin wave wavefront, \textcolor{black}{in the continuous-wave regime of spin-wave propagation and in the stationary state}, without needing any optical reference path. By selecting the applied field, frequency, and detection phase, the spin wave images can be made sensitive to the precession amplitude and phase. Our results demonstrate that infrared optical strobe light can serve as a versatile platform for magneto-optical probing of magnetization dynamics, with potential implications in investigating hybrid magnonic systems.  

\end{abstract}

\flushbottom
\maketitle

\thispagestyle{empty}

\section{Introduction}

Hybrid magnonic systems are rising contenders for quantum information transduction owing to their capability of coherently connecting distinct physical platforms in quantum systems \cite{awschalom2021quantum}. Recent studies have revealed strong and coherent hybridization of magnons with phonons, microwave photons, and optical light, with the observation of characteristic phenomena that further give rise to emerging quantum engineering functionalities \cite{lachance2019hybrid,li2020hybrid,yuan2022quantum,bhoi2019photon,li2021advances,chumak2022advances}.  

Due to the increasing demand of chip-integrable circuit elements hosting such hybrid functionalities, the past decade has witnessed rapid developments in film-based hybrid magnonic systems with complementary micro- and nano-structuring capabilities such as using photo- and e-beam lithography \cite{flebus20242024,hou2019strong,li2019strong,shao2021roadmap}. In accordance with this advancement, an important task is the sensitive and reliable detection of spin dynamics of the core magnonic device components, which usually involve coupled magnetic multilayers consisting of ferromagnetic (FM) metals, semiconductors, and dielectrics, such as yttrium iron garnet (Y$_3$Fe$_5$O$_{12}$, YIG) \cite{serga2010yig}-based heterostructures \cite{liu2024strong,li2020coherent,klingler2018spin,chen2018strong,qin2018exchange,qin2021nanoscale,vilsmeier2024spatial,santos2023magnon}. 

Such a task renders the optical techniques very appealing due to their highly localized probe and potential spatial-resolving capability, compared to conventional electrical or transmission measurements \cite{makiuchi2024persistent,li2021phase,trossman2019phase,xiong2024combinatorial,xiong2024hybrid}. In addition, as both the amplitude and phase emerge as relevant state variables in quantum systems, the ability to track the magnon phase relative to other excitations, such as another driving microwave photon, phonon, or magnon, in a hybrid magnetoelectric circuit becomes paramount. 

In such a context, an `optical stroboscope' probing magneto-optical effect of the film samples emerges naturally as a neat technique with combined spatial- and phase-resolving capabilities \cite{nembach2013mode,yoon2016phase}. The stroboscopic effect is a phenomenon caused by aliasing when a continuous, cyclic motion is represented by a series of short or instantaneous samples (as opposed to a continuous view) at a sampling rate close to the period of the motion. Thus, to probe spin dynamics that are usually in the gigahertz (GHz) regime, the strobe light has to be modulated as fast as the ferromagnetic resonance (FMR). However, such a requirement is incompatible with most popular spectroscopic wavelengths (usually in or near the UV-VIS range) \cite{tanksalvala2024element}. As a result, pump-probe technique employing pulsed lasers \cite{kirilyuk2010ultrafast,qin2021nanoscale,dreyer2022imaging}, \textcolor{black}{Brillouin (inelastic) light scattering (BLS) leveraging photon Stokes shifts \cite{sebastian2015micro,lendinez2023nonlinear,wang2023deeply,grachev2023reconfigurable,grachev2022strain,sadovnikov2022reconfigurable},} and nitrogen-vacancy(NV) magnetometer using proximal spin-dipole field interactions \cite{andrich2017long,zhou2024sensing,bertelli2020magnetic} have been adopted for investigating spin dynamics. Nevertheless, augmenting the phase-resolving capability requires additional nontrivial hardware implementation \cite{aleman2020frequency,freeman2020brillouin,l2023correlation,wojewoda2023phase,hashimoto2018phase}. For example, an auxiliary reference light path (with a constant phase) needs to be introduced to interfere with the scattered light in a phase-resolved BLS setup \cite{serga2006phase,fallarino2013propagation}.    

Thanks to the mature telecom technology, the output of a telecom fiber laser at 1550-nm wavelength in the infrared (IR) band can be fully modulated using an electro-optical intensity modulator (EOM) \cite{xiong2020experimental,xiong2024magnon} at the GHz range, thus can be used as a strobe light probe for spin dynamics. \textcolor{black}{Compared to conventional pump-probe and BLS techniques, the IR strobe light probe presents several unique properties}: 

\textcolor{black}{(1) the strobe feature allows to accurately trace the spin dynamics with explicitly-defined phase contributions (tracking);} 

\textcolor{black}{(2) the detected spin precessional phase and amplitude can be used to directly and concurrently construct the spin wave's wavefront and intensity, in the continuous-wave (cw) regime of spin wave propagation and in the stationary state, just from the different channels of the lock-in amplifier (mapping);} 

\textcolor{black}{(3) the resonant frequency of each magnon mode is directly obtained by the spectroscopy, without the need for any temporal-spectral transformation as often encountered in pump-probe techniques (dispersion);} 

\textcolor{black}{(4) due to the phase accumulation arising from the spin wave propagation, the spin wave group velocity for each mode can be directly extracted (propagation); and}

\textcolor{black}{(5) unlike the UV-VIS, the IR light is nearly perfectly transparent to YIG and many rare-earth-doped YIG derivative materials -- the same reason YIG and its doped-counterparts are excellent telecom magneto-optical devices (transmissivity) \cite{onbasli2016optical,fakhrul2019magneto,dash2023surface,dash2024band}.} 

In particular, with respect to hybrid magnonics, the exceptional transmission characteristic makes the IR wavelength nicely suited for magneto-optical probing of spin dynamics of respective layers and their relative precessional phase in FM-metal/YIG heterostructures, employing concurrent Kerr and Faraday effects. Such a concept has been previously demonstrated in spectroscopic measurements of YIG/Permalloy(Py), revealing the coherent magnon-magnon hybridization between the uniform mode of Py and the perpendicular standing spin wave modes of YIG \cite{xiong2022tunable,xiong2020probing, xiong2020detecting,inman2022hybrid}. However, the ability of such a technique to perform phase-resolved spatial spin-wave imaging has remained elusive.  

In this work, we further consider the spatial- and phase-resolved spin wave imaging using 1550-nm strobe light. Previously, the spectroscopic study has focused on the perpendicular standing spin waves (PSSWs) of YIG and the coherent interaction with the uniform mode of Py \cite{xiong2020detecting,xiong2020probing, xiong2022tunable}. To demonstrate the spatial mapping capability of the technique, we herein focus on the backward volume spin wave (BWVSW) modes with a magnetic underlayer of Py that serves as a retroreflective mirror. Compared to the surface spin waves (e.g. in the Damon-Eshbach \cite{eshbach1960surface,bhaskar2020backward}) geometries, the use of bulk traveling waves allows to exemplify the concurrent detection using magneto-optical Kerr and Faraday effects. 

\section{Experimental Setup}

\begin{figure}[htb]
 \centering
 \includegraphics[width=3.2 in]{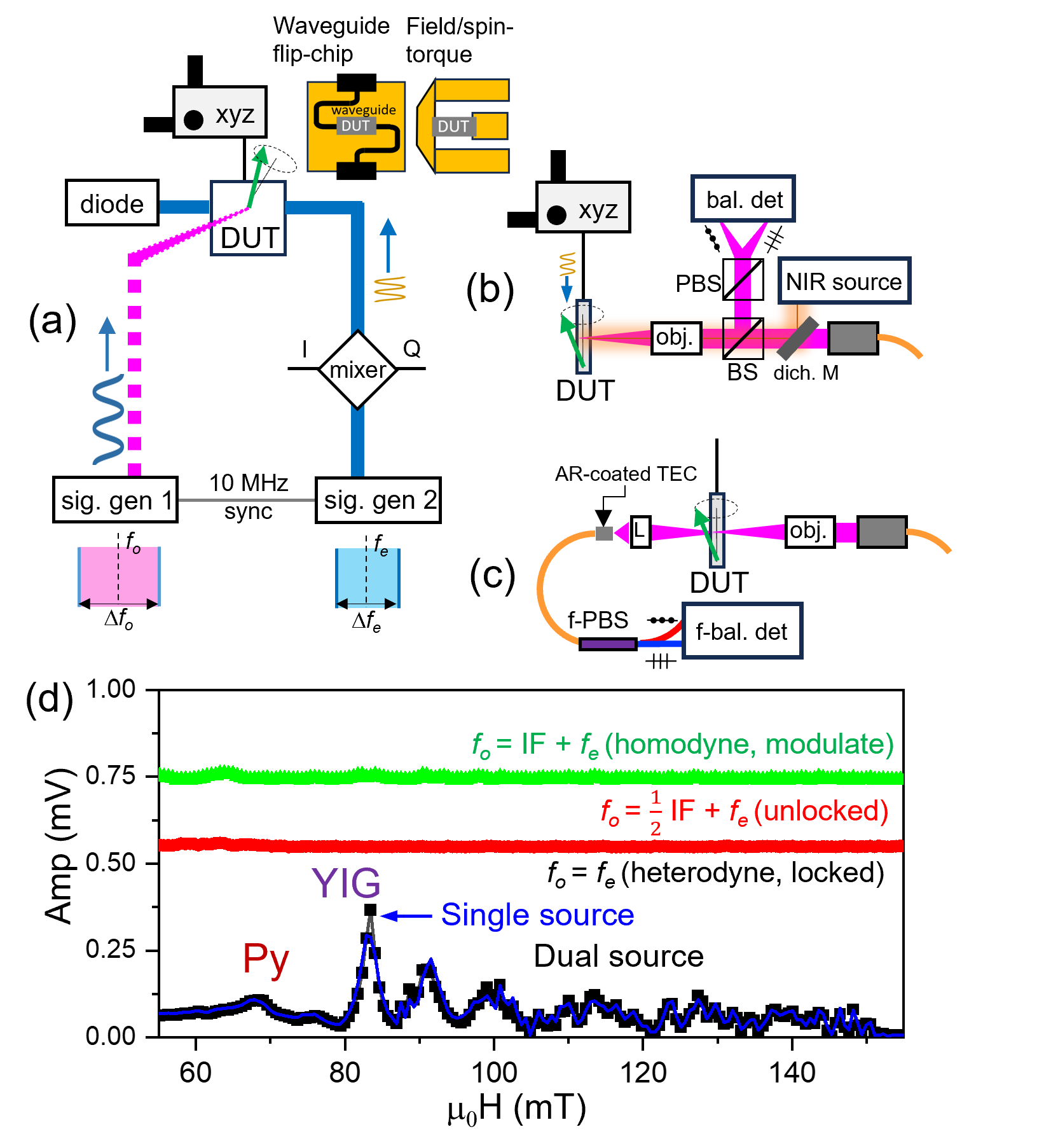}
 \caption{(a) Schematic illustration of the strobe light and microwave transmission measurement setup. Two signal generators (sig. gen 1 and 2) can be synchronized. One modulates the strobe light and probes the spin dynamics optically at $f_o$, while the other drives the DUT at $f_e+$IF and detects the spin dynamics by common transmission measurement using a power diode (heterodyne case). \textcolor{black}{The microwave signal can be either sent through a stripline or waveguide that couples to the magnetic film sample in a ``flip-chip'' measurement configuration, or directly into the sample (often metallic) that excites the FMR by means of the local rf field and/or spin-torque mechanisms.} (b) The reflective geometry: strobe light of 45$^\circ$ polarization state passes through an objective lens (obj.), focused atop the sample (DUT), and the modulated polarization states (due to magneto-optical effects) of the retro-reflected light is analyzed by a PBS and a balanced photo-detector (bal.det). Illumination can be made by adding a dichroic mirror (dich.M) and a near-infrared (NIR) light source. (c) The transmission geometry: the focused light passes through the sample, re-collimated by an aspheric lens (L), and recollected by a AR-coated, TEC fiber ferrule. The subsquent light analysis can be done in all-fiber format using a fiber(f)-PBS and fiber-port balanced photo-detector (f-bal.det). (d) Representative signal traces measured for a Py/YIG bilayer using concurrent magneto-optical Kerr(Py) and Faraday(YIG) effects: with the IF, heterodyne case ($f_o=f_e$) using two locked microwave source (symbol) and a single source (line) with a rf splitter, homodyne case ($f_o=f_e+$IF) so that the same frequency at DUT, and unlocked case ($f_o = f_e+\frac{\mathrm{IF}}{2}$). } 
 \label{fig:Fig1}
\end{figure}

Despite being an optical technique, one neat feature of using 1550-nm strobe light lies in the direct, facile setup integration to almost any standard microwave transmission measurements. The solid-line path in Fig. \ref{fig:Fig1}(a) exemplifies common microwave transmission measurements, such as vector-network-analyzer (VNA) measurement, direct power diode absorption spectroscopy, and field-modulation FMR, with the basic principle of detecting the absorption of the input microwave energy caused by resonant excitation of the magnetic material, i.e. the device-under-test (DUT). \textcolor{black}{Depending on the different FMR excitation schemes, the microwave signal is either sent through a stripline or waveguide that couples to the magnetic film sample in a ``flip-chip'' measurement configuration, or directly into the sample (often metallic) that excites the FMR by means of the local rf field and/or spin-torque mechanisms \cite{shao2021roadmap,chumak2022advances}.}

By simply branching out a portion of the microwave signal for electro-optical modulation, a strobe light probe by means of magneto-optical Kerr or Faraday effects can be introduced, as indicated by the dotted-line in Fig.\ref{fig:Fig1}(a). Such a microwave diffluence can be realized by using a rf-splitter from a single rf source, where the microwave excitation and optical detection coincide at the same frequency, representing a `cw-stroboscope'. Alternatively, such a function can be achieved with two rf generators that are synchronized using the 10 MHz common reference (phase-locked). In such a case, the microwave excitation (at $f_e$) and optical detection (at $f_o$) can be made to occur at different frequencies for detecting harmonic resonances, representing a `pump-probe stroboscope'. 

Depending on the DUT's magneto-optical characteristic and measurement goal, the optical path can take the form of either reflective (terminal) or transmission (in-line) geometries:

\textit{(i)} For the reflective version, the sample is treated similarly to a `Faraday mirror'. As shown in Fig.\ref{fig:Fig1}(b), the intensity-modulated strobe light was initially set to a linear polarization state of 45$^\circ$ using a fiber polarizer (polarization balanced state), then passed through a collimator, converting the fiber light to a real-space beam. The laser spot is focused down to $\sim$10 $\mu$m atop the sample surface using an IR objective lens. The magnetic sample was in-plane (IP) magnetized in the static situation, but the dynamic out-of-plane (OOP) polarization component caused by the spin dynamics can be sensitively detected, in the form of a small polarization perturbation away from the 45$^\circ$ balanced state. The retro-reflected light beam was analyzed using a polarizing beam splitter (PBS) and a balanced photo-detector, and then sent to a lock-in amplifier for signal demodulation. Lastly, a dichroic mirror and an NIR light source can be inserted to facilitate sample imaging and visualization combined with an IR-sensitive camera.      

\textit{(ii)} For the transmission version, the sample is used similarly to an in-line `Faraday rotator'. As shown in Fig. \ref{fig:Fig1}(c), one can leverage extensive fiber-optical components and further minimize the use of real-space optics. As shown in Fig.\ref{fig:Fig1}(c), after passing through the sample, the transmitted light can be coupled, using an aspheric lens, into an FC-terminated, thermally-expanded-core (TEC) fiber ferrule with special anti-reflective (AR) coating. The coupled light was then sent to a fiber-PBS and analyzed using a fiber-port balanced detector. In this case, the use of real-space optics consists of only two focusing lenses in the vicinity of the sample surfaces. Such a geometry makes sense when the spin dynamics of the DUT is used for in-line optical modulation (insertable DUT). For plano-plano sample geometries, e.g., magnetic films and multilayers, such a configuration can be facile installed on a pre-aligned fiber U-bench or other similar setup.   

Along the electric path, a heterodyne technique can be activated by adding an I-Q (in-phase and quadrature) mixer that mixes the microwave driving frequency, $f_e$, with another intermediate frequency, IF (100 kHz in this work) to create and control the sidebands. By sending two sinusoidal signals of IF with a constant phase to the respective I and Q ports, a single sideband of $f_e+$IF can be picked out and used for (electrically) driving the spin dynamics of the sample. All other harmonics, such as the lower sideband, $f_e-$IF, can be suppressed by optimizing the I-Q channels input phase. In such an arrangement, the strobe light (modulated at $f_o$) arrives slightly later in the precession cycle and over time, the demodulated magneto-optical response traces out complete spin precession cycles. Figure \ref{fig:Fig1}(d) shows an example signal trace of a YIG disc(350-$\mu$m)/Py(50-nm) bilayer at 4-GHz, using the heterodyne detection mechanism with two synchronized microwave sources (symbol). The trace is nearly identical with that using a single source and a rf splitter (line), or using harmonic frequencies, i.e. $f_e = n \times f_o$, $n$ is an integer. When the strobe light frequency is the same as the DUT driving frequency ($f_o = f_e+$IF) representing the homodyne case, a small modulation of the signal can be observed but is much less pronounced than the heterodyne case. At other frequencies, such as ($f_o = f_e+\frac{\mathrm{IF}}{2}$), null signal is detected corresponding to an example unlocked case.

\section{Results and Discussion}

\subsection{Spectroscopy}
The spectroscopy part of the technique has been outlined in earlier reports \cite{xiong2020detecting,xiong2020probing,xiong2022tunable,inman2022hybrid,li2019simultaneous}. The essence of the detected signal after demodulation by the lock-in amplifier captures the total phase accumulation in the experiment, where the lock-in amplifier's in-phase ($X$) and quadrature ($Y$) channels read: $X \propto \delta m_z P_0 \textrm{cos}(\phi_{eo} - \phi_m)$ and $Y \propto \delta m_z P_0 \textrm{sin}(\phi_{eo} - \phi_m)$. The signal is proportional to the amplitude of the film-normal component of the oscillating magnetization ($\delta m_z$) and the laser power ($P_0$), but the key asset lies the enclosed phase information: 

(1) $\phi_{eo}$ represents the instrumental, `magnetic-field-independent' phase induced by the optical and the electric paths difference. It is a character determined by the setup (optical delay, microwave and fiber cable lengths, etc) but can be controlled to accommodate tailored samples and serve as a self-calibrated reference to the magnetic phase of interest \cite{xiong2020detecting}. 

(2) $\phi_{m}$ is the magnetic phase, which is the signal of interest due to magneto-optical effects, and can originate from spin-wave propagation \cite{shiota2020imaging}, magnon-magnon interaction \cite{xiong2020probing,xiong2022tunable}, and spin-orbit torques \cite{li2019simultaneous,shiota2021inhomogeneous}. 

A quintessential feature of strobe light spin wave probing lies in the strong and robust phase correlation of the detected magnons to the external microwave drive. To demonstrate such an attribute, we measured a YIG disc(350-$\mu$m)/Py(50-nm) bilayers in which the Py layer is either a continuous film or a patterned structure using photolithography. 

\begin{figure}[htb]
 \centering
 \includegraphics[width=3.2 in]{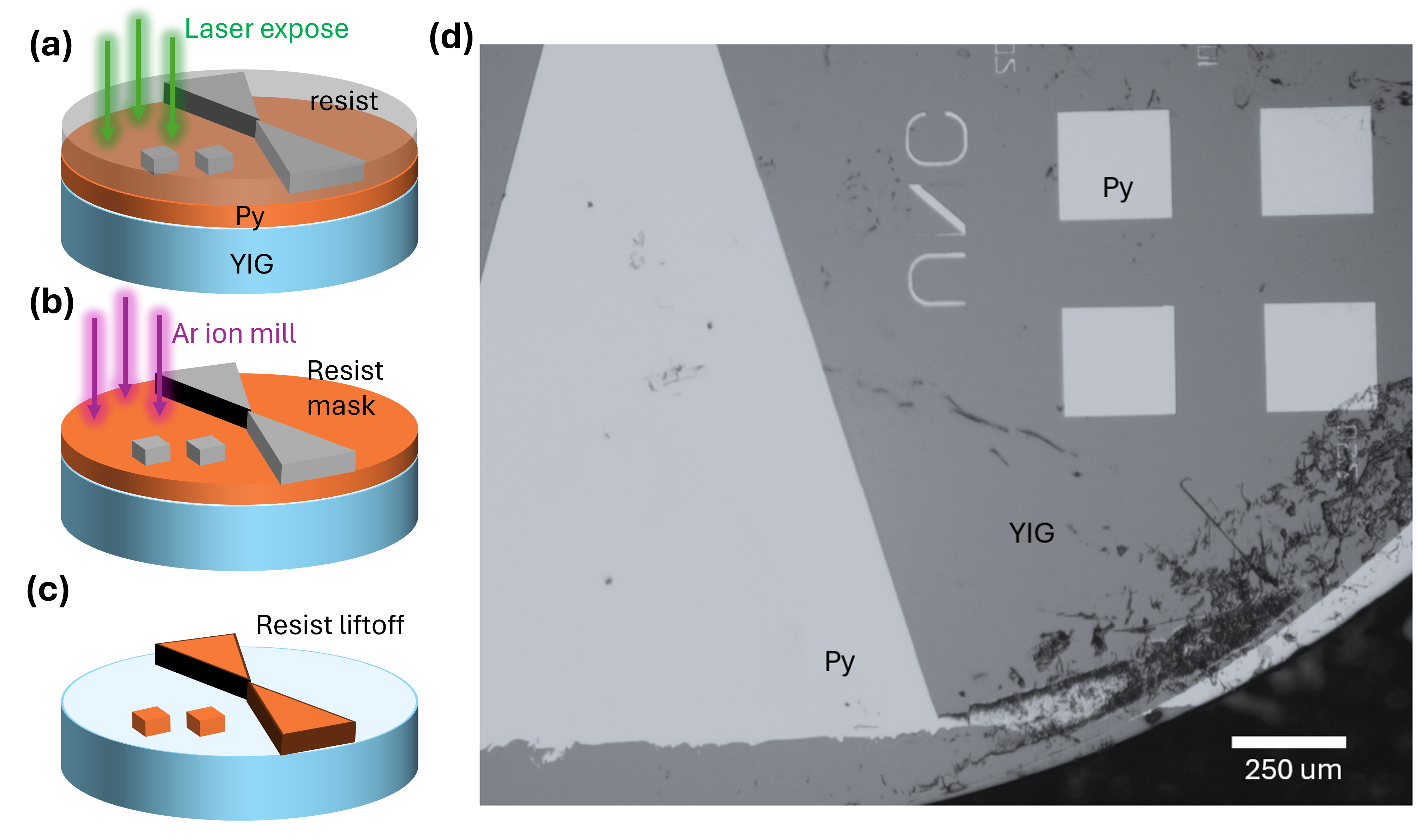}
 \caption{(a,b,c) Illustration of the sample fabrication process. (a) The Py layer is deposited by magnetron sputtering on a double-side-polished commercial YIG substrate. A layer of photoresist is spin coated and the resist mask pattern is defined by direct laser writing. (b) The pattern was transferred to the Py layer by Argon ion milling in a high-vacuum chamber through the resist mask down to the YIG surface. (c) The final liftoff process to expose the patterned Py microstructures. (d) Photograph image showing the fabricated square dot array in the vicinity of a large triangular bowtie (area of interest). The scale bar is 250 $\mu$m. }
 \label{fig:Fig1.5}
\end{figure}

\textcolor{black}{The sample fabrication process is illustrated in Fig.\ref{fig:Fig1.5}. The YIG substrate is a 350-$\mu$m double-side polished commercial disc and the Py layer is deposited by magnetron sputtering. For fabricating Py microstructures, the resist mask pattern was defined by photoresist coating and direct laser writing, Fig.\ref{fig:Fig1.5}(a). The pattern was then transferred to the Py layer by Argon ion milling in a high-vacuum chamber through the resist mask down to the YIG surface, Fig. \ref{fig:Fig1.5}(b), and finally a resist liftoff step to expose the Py microstructure, Fig.\ref{fig:Fig1.5}(c). Figure \ref{fig:Fig1.5}(d) is a photograph image showing the fabricated $2 \times 2$ square dot array in the vicinity of a large triangular bowtie, whose spin wave imaging will be discussed later.} 

\begin{figure}[htb]
 \centering
 \includegraphics[width=3.5 in]{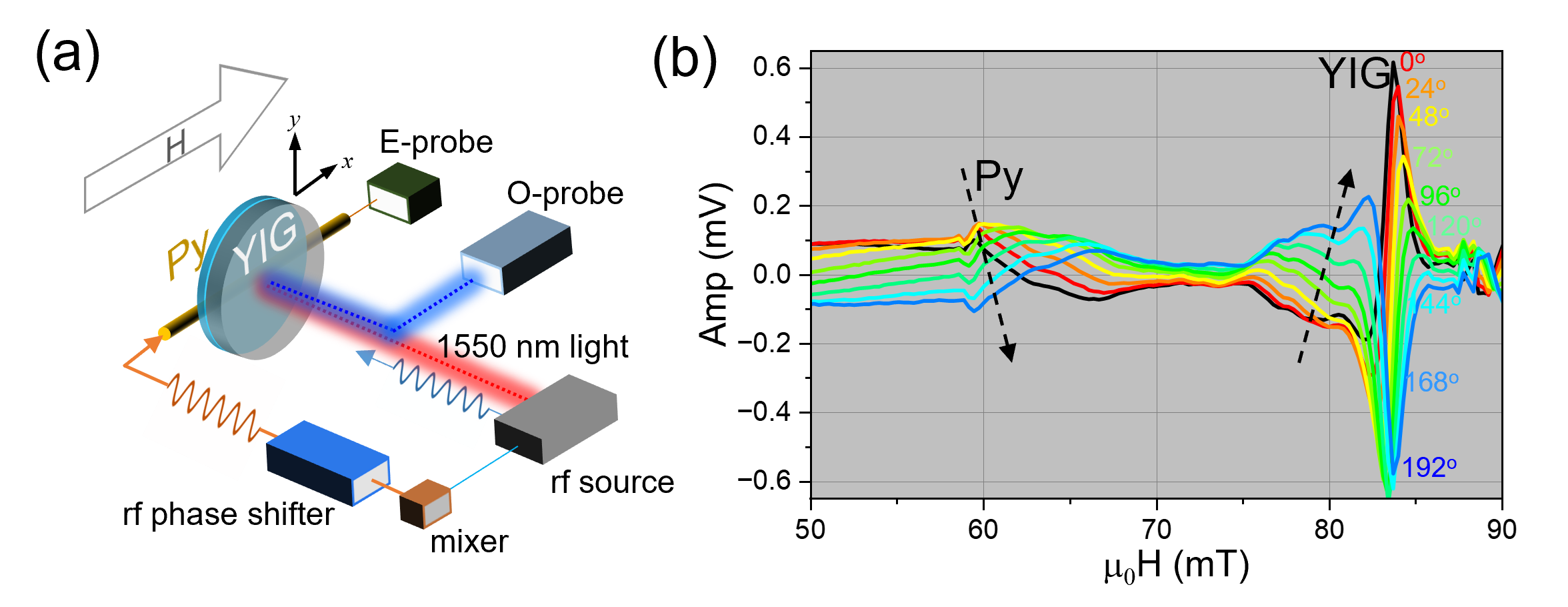}
 \caption{(a) Experimental setup of the spin-wave spectroscopy and imaging (O-probe) with complementary field-modulation FMR using a rf power diode (E-probe). The sample is a Py/YIG continuous-film bilayer: the YIG is a 350-$\mu$m double-side polished commercial disc and the Py is 50-nm deposited atop. (b) Strobe light detected signal at 4-GHz with varying microwave phase tuned by a rf phase shifter: $\phi_{rf}$ from 0$^\circ$ to 192$^\circ$ at an increment of 24$^\circ$. }
 \label{fig:Fig2}
\end{figure}

\begin{figure*}[htb]
 \centering
 \includegraphics[width= 5.4 in]{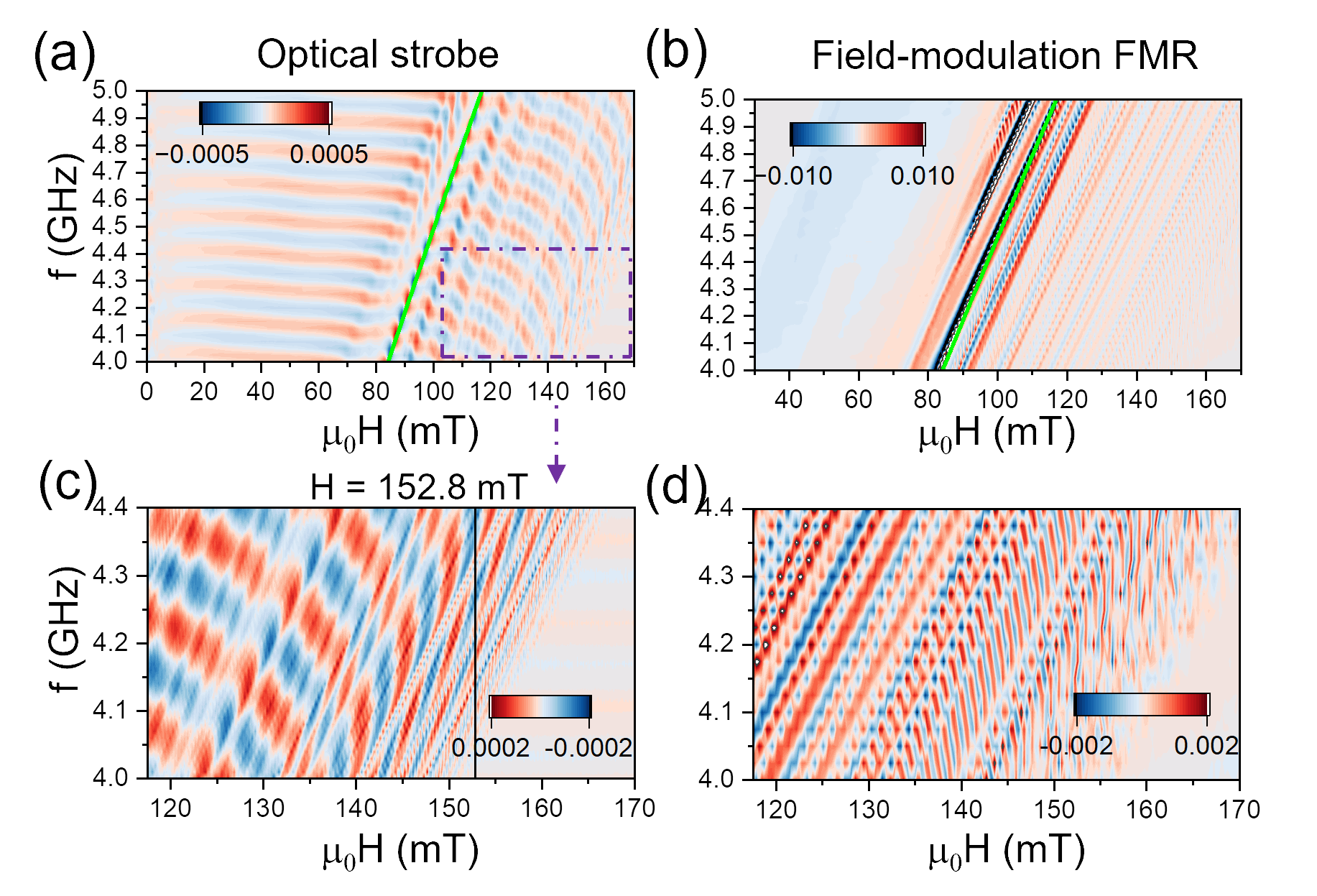}
 \caption{(a,b) Comparison of $f-H$ spectra detected by (a) strobe light (local) and (b) the complementary field modulation FMR measurement (global). \textcolor{black}{The Kittel mode is highlighted by the solid-line. Fitting to the Kittel mode yields an $M_s$ value of $\sim$ 0.177 T using an $\gamma$ value of 27 GHz/T}. The zoom-in regime of the respective (c) optical, and (d) field modulation spectra with finer scan steps. Example extraction of dispersion by means of imaging analysis will be demonstrated later at $H$ = 152.8 mT. }
 \label{fig:Fig3}
\end{figure*}

Figure \ref{fig:Fig2}(a) illustrates our experimental setup for the spectroscopy and imaging: the sample is measured under the reflective geometry with an in-plane bias magnetic field along $x$. The phase of the electric path relative to the strobe light path, in addition to the instrumental $\phi_{eo}$, is further tuned by a rf phase shifter (adding an additional $\phi_{rf}$). Figure \ref{fig:Fig2}(b) shows the detected signal at 4-GHz under selective $\phi_{rf}$ values, from 0$^\circ$ to 192$^\circ$. 

Both the Py and YIG resonance profiles can be modeled by a complex Lorentzian lineshape consisting of the symmetric $S(H)$ and antisymmetric $A(H)$ functions: $S(H)=\frac{\Delta H^2}{[H - H_{res}]^2+\Delta H^2}$, and $A(H)=\frac{H-H_{res} \Delta H}{[H - H_{res}]^2+\Delta H^2}$, where $H_{res}$ is the resonance frequency and $\Delta H$ is the resonance linewidth. As shown in Fig.\ref{fig:Fig2}(b), at this frequency, the precessional phase of Py changes in an opposite fashion with that of the YIG, upon varying the detection phase $\phi_{rf}$.    

Comprehensive $f-H$ dispersion spectra were measured using simultaneous strobe light probe (local), Fig.\ref{fig:Fig3}(a) and complementary field-modulation FMR (global), Fig.\ref{fig:Fig3}(b). Given the measurement geometry and the spin wave dispersion, we confirm the measured BWVSW modes in the 350-$\mu$m YIG sample. The BWVSWs are excited through coupling of the OOP component of the alternating rf field with the OOP component of the dynamic magnetization. Typically, the excitation of the uniform Kittel mode overwhelms that for the spin wave modes, therefore, the FMR signal is often at least one order of magnitude larger in a conventional microwave transmission measurement. Here, by using the strobe light probe, Fig.\ref{fig:Fig3}(a), the sensitivity is notably enhanced for the spin wave modes referenced to that of the FMR, in contrast to the field modulation method in Fig.\ref{fig:Fig3}(b), where the strong FMR renders the spin wave traces hardly discernible.  

Notably, in Fig.\ref{fig:Fig3}(a) and (c), a periodically alternating phase evolution of the optical strobe signal can be clearly observed. This spectroscopic character, as shown by our subsequent analysis, bestows information of the propagating wavefront and allows the direct extraction of the spin wave group velocity $v_g$. Using a constant-field slice of the spectra ($H$ = 140.0 mT), the frequencies of each individual spin wave at that field can be taken and the $(f,H)$ points can be converted to the wavevector using the dispersion relation. This allows for the calculation of the group velocity $v_g$.  The evolution is nearly independent of magnetic field on the low field side (left of the strong Kittel mode) as shown in Fig.\ref{fig:Fig3}(a). However, on the high field side (right of the Kittel mode), as the field increases, the mode number increases and the phase evolution gradually bends towards lower frequencies. The magneto-optical signals exhibit a clear cutoff following the same $f-H$ slope as the Kittel mode. The observation of such cutoff behavior is another elemental character of the BWVSW, in which a finite wavevector($\textbf{k}$) can lower the magnon frequency from the Kittel (zero-$\textbf{k}$) mode, in other words, increase the resonance field at a fixed frequency. On the other hand, such a feature is absent in the complementary field modulation FMR measurement, Fig.\ref{fig:Fig3}(b) and (d), which is insensitive to the spin precession phase.

\begin{figure}[htb]
 \centering
 \includegraphics[width=3 in]{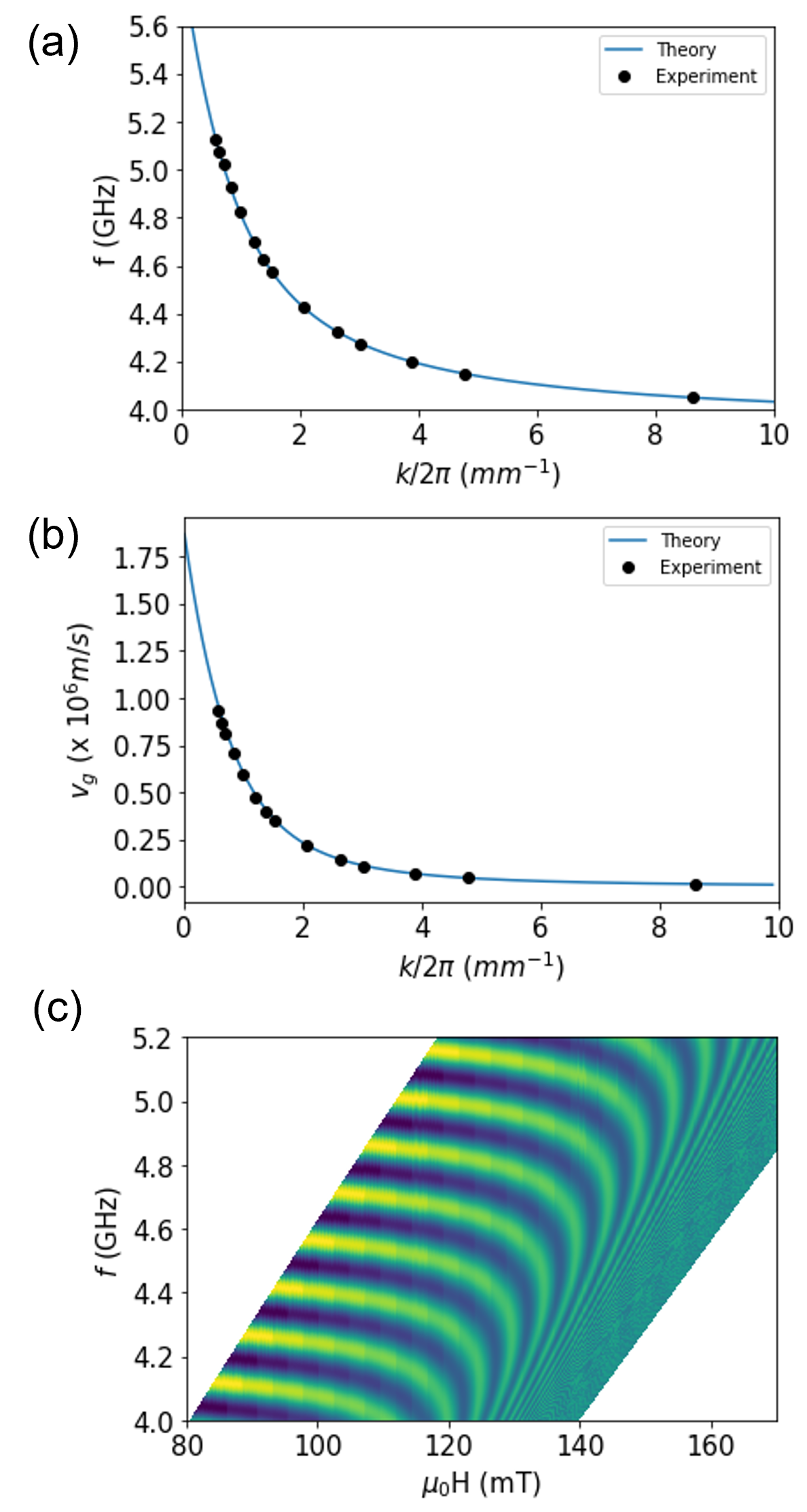}
 \caption{(a) Dispersion curve of BWVSW, $f=\omega_{BV}/2\pi$ as a function of $k$, at $H=140$ mT calculated from Eq. (\ref{eq02}) with points extracted from the $f-H$ dispersion spectra. (b) The corresponding $-v_g$ calculated from the dispersion. (c) Calculated evolution of $\phi_{}$ from Eq. (\ref{eq01}) taking $L=2.0$ m, $d_p=0.2$ mm, and $d=350$ $\mu$m.}
 \label{fig:Fig4}
\end{figure}

To analyze the above phase evolution, we formulate the measured phase as:

\begin{equation}
    \phi_{}=\omega L/c + \omega d_p/v_g
\label{eq01}
\end{equation}
The first term corresponds to $\phi_{eo}$, with the microwave propagating close to the speed of light ($c=3\times 10^8$ m/s) along an effective path difference, $L$ \cite{li2019simultaneous}. The second term corresponds to the phase delay due to spin wave propagation from the `point of excitation', i.e. the coplanar waveguide (CPW) to the `measurement location', i.e. the laser spot, with an effective traveling distance of $d_p$ and a group velocity of $v_g$. For the first part, $L$ can be determined from the frequency-dependent phase evolution when the magnetic field is much smaller than the Kittel mode's resonance field (e.g $H\sim 40.0$ mT), yielding $L \sim 2.0$ m. The second part is magnetic-field-dependent, as the value of $v_g$ varies with the magnetic field. We can calculate $v_g$ from the BWVSW dispersion relationship, with $v_g=\partial\omega_{BV}/\partial k$ \cite{serga2010magnonics}, and:

\begin{equation}
    \omega_{BV}=\gamma\sqrt{H\left(H+M_s {1-\exp{(-kd)} \over kd}\right)}
\label{eq02}
\end{equation}
where $\gamma=2\pi\times 28$ GHz/T is the gyromagnetic ratio of YIG taking the $g$ factor as 2, $M_s=0.175$ T is the magnetization of YIG, and $d=350$ $\mu$m is the YIG thickness. Figure \ref{fig:Fig4}(a) shows the calculated $\omega_{BV}$ as a function of $k$ at $H=140.0$ mT, along with the individual points extracted from the $f-H$ stroboscopic spectra. $\omega_{BV}$ extends by around 1.7 GHz, which matches well with the observation in Fig. \ref{fig:Fig3}(a). Figure \ref{fig:Fig4}(b) shows the calculated $v_g$ using $v_g=\partial\omega_{BV}/\partial k$ along with the $v_g$ calculated from the points in Fig. \ref{fig:Fig4}(a). The points from the spectra show very good agreement with the theoretical model. 

Figure \ref{fig:Fig4}(c) shows the theoretical plot of $\phi$ from Eqs. (\ref{eq01}) and (\ref{eq02}) using $d_p=0.2$ mm. \textcolor{black}{From the low-field side, the Kittel mode ($\bf{k}$ = 0) couples most efficiently to the rf antenna. As field increases, the magnon wavelengths become smaller and the coupling becomes less effective, thus reducing the excitation amplitude, along with the $\phi$ modulation caused by the spin wave propagation. The dispersion cutoff on the right-hand side (where the phase evolution also diverges) corresponds to $k=\infty$ as the low-frequency bound of the BWVSW mode. The calculated $\phi$ evolution shows a good agreement with} the experimental $f-H$ contour plot in Fig. \ref{fig:Fig3}(a).

\subsection{Imaging}

Using the strobe light probe, the detected precessional phase contrast can be directly used to construct the spin-wave wavefront \textcolor{black}{in the stationary state}, without needing any additional optical reference paths. By using a long working-distance IR-band objective lens combined with a precise 3-D micro-positioner ($x,y,z$), we demonstrate the capability of spin-wave spatial mapping in our YIG/Py samples.

\subsubsection{Field-dependent Wavefront}

\begin{figure}[htb]
 \centering
 \includegraphics[width=3.3 in]{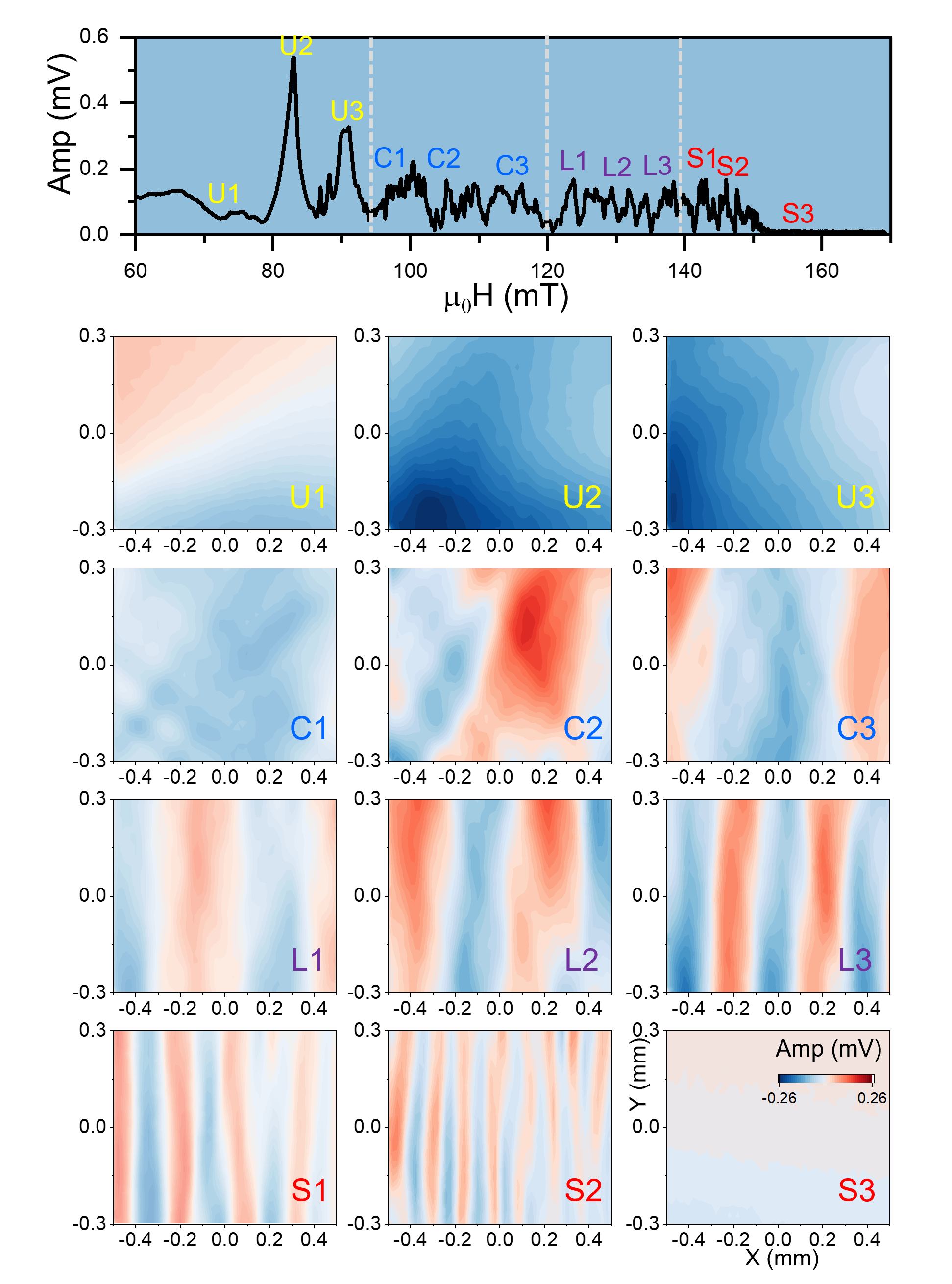}
 \caption{Top panel: A 1-D field-scan amplitude trace measured at 4-GHz for the Py/YIG bilayer. As the field increases, the resonance is dominated progressively by the uniform (U), caustic (C), long (L) and short (S) wavelength regimes. In each regime, the 2-D wavefront maps at three representative magnetic field values were scanned. U1: 71.9, U2: 83.5, U3: 89.2. C1: 95.0, C2: 106.6, C3: 118.1. L1: 126.8, L2: 132.6, L3: 138.4. S1: 144.2, S2: 150.0, S3: 158.6. (Unit in mT) }
 \label{fig:Fig4_1}
\end{figure}

First, we focus on the continuous Py/YIG film. The Py/YIG sample is chip-flipped atop a broadband CPW. We performed 2-D scans (along $x-y$) in the vicinity of the CPW. Figure \ref{fig:Fig4_1} shows the representative 2-D wavefront maps at selective magnetic fields. The top panel shows the 1-D field-scan amplitude at 4-GHz near the YIG resonance regime. We studied the field-dependent wavefront maps by dividing the field-scan trace into four distinct sections (U, C, L, S). 
 
The uniform (U) regime covers from the YIG's Kittel mode ($\sim 75.0$ mT) up to $H=95.0$ mT. Three maps were scanned at fields: U1 (prior to YIG FMR): $H=71.9$ mT, U2 (on YIG FMR): $H=83.5$ mT,  and U3: $H=89.2$ mT. Quasi-uniform spin precession was indicated throughout this regime due to the collective in-phase motion of the magnetization.   

As the field increases, dipolar spin-wave modes start to emerge with increasing wavevector($\textbf{k}$). According to earlier reports, the spin dynamics commonly enter a caustic (C) regime \cite{schneider2010nondiffractive,sebastian2013nonlinear,swyt2024magnonic,muralidhar2021femtosecond,temdie2024probing,martyshkin2024nonreciprocal}, in which a well-defined propagation direction (group velocity) is favored due to anisotropies in the YIG dispersion relation. \textcolor{black}{Such an effect is characterized by the `slowness curve', i.e., a curve winding around the origin in wavevector space that indicates the slowness (reciprocal of velocity) at different wavefront angles \cite{wartelle2023caustic,bertelli2020magnetic,bertelli2021imaging,sebastian2013nonlinear,martyshkin2024nonreciprocal,makartsou2024spin}. The inflection point on such a curve corresponds to the favored group velocity direction, which typically occurs around the unity of $kd$ where $d$ is the thickness of the film \cite{wartelle2023caustic}.} Moreover, in larger scale films such as the present case, multiple scattering sources arising from the edges, film defects, or magnetic underlayers can result in spin-wave interference, creating characteristic Talbot-like diffraction patterns \cite{makartsou2024spin}. We observed such a phenomenon in our scanned 2-D wavefront map at selective magnetic fields: C1: $H=95.0$ mT, the spin precession is largely uniform and anti-phase to the microwave drive with the observation of caustic nodes. C2: $H=106.6$ mT, long-wavelength wavefronts start to emerge (color contrast) with superimposed caustic nodes, and C3: $H=118.1$ mT, the transition between the caustic regime to the long-wavelength regime. By performing a 2-D Fast-Fourier-Transformation (FFT), we estimated the range of the caustic angle is between 112.4$^\circ$ to 122.1$^\circ$, \textcolor{black}{in good agreement with previously reported values \cite{wartelle2023caustic,bertelli2020magnetic,bertelli2021imaging,sebastian2013nonlinear,martyshkin2024nonreciprocal,makartsou2024spin}.} However, due to other convoluting mechanisms, a detailed investigation of such diffraction patterns may merit a separate study.

Next, the dynamics enters the well-defined long-wavelength regime, characterized by clearly identified BWVSW wavefronts propagating along the $x$ direction. We show wavefront maps at selective fields in two regimes in which the wavelength is greater (L) or smaller (S) than the thickness of the film ($d=350$ $\mu$m). We show example 2-D wavefront maps at L1: $H=126.8$ mT, L2: $H=132.6$ mT,  and L3: $H=138.4$ mT. In the L-regime, the intensity of the spin waves maintains roughly a stable amplitude. Further, at the transition regime from the C to L ($\sim 120.0$ mT), the dispersion exhibits a gap, which is likely due to the spin-wave stopband caused by the Py underlayer according to earlier reports \cite{riedel2023hybridization,vilsmeier2024spatial}. In the S-regime, we show maps at S1: $H=144.2$ mT and S2: $H=150.0$ mT. The intensity of the spin wave amplitude gradually decreases as the field further increases, and finally diminishes, e.g. at S3: $H=158.6$ mT, due to the cutoff behavior (high-$\textbf{k}$ saturation) in the BWVSW dispersion relation \cite{serga2010yig}.  

\subsubsection{Wavevector}

The scanned 2-D maps further allow the extraction of the wavelengths and wavevectors via FFT. In BLS measurement, the spin-wave wavevector can be obtained by varying the angle of incidence of the laser beam \cite{sebastian2015micro}, and the result only carries the information relevant to the single focused laser spot. Using the 2-D scanned wavefront maps with larger dimensions along both $x$ and $y$ (up to mm in this work), the obtained information entails all the contributing wavevectors and their distribution. 

\begin{figure}[htb]
 \centering
 \includegraphics[width=3.4 in]{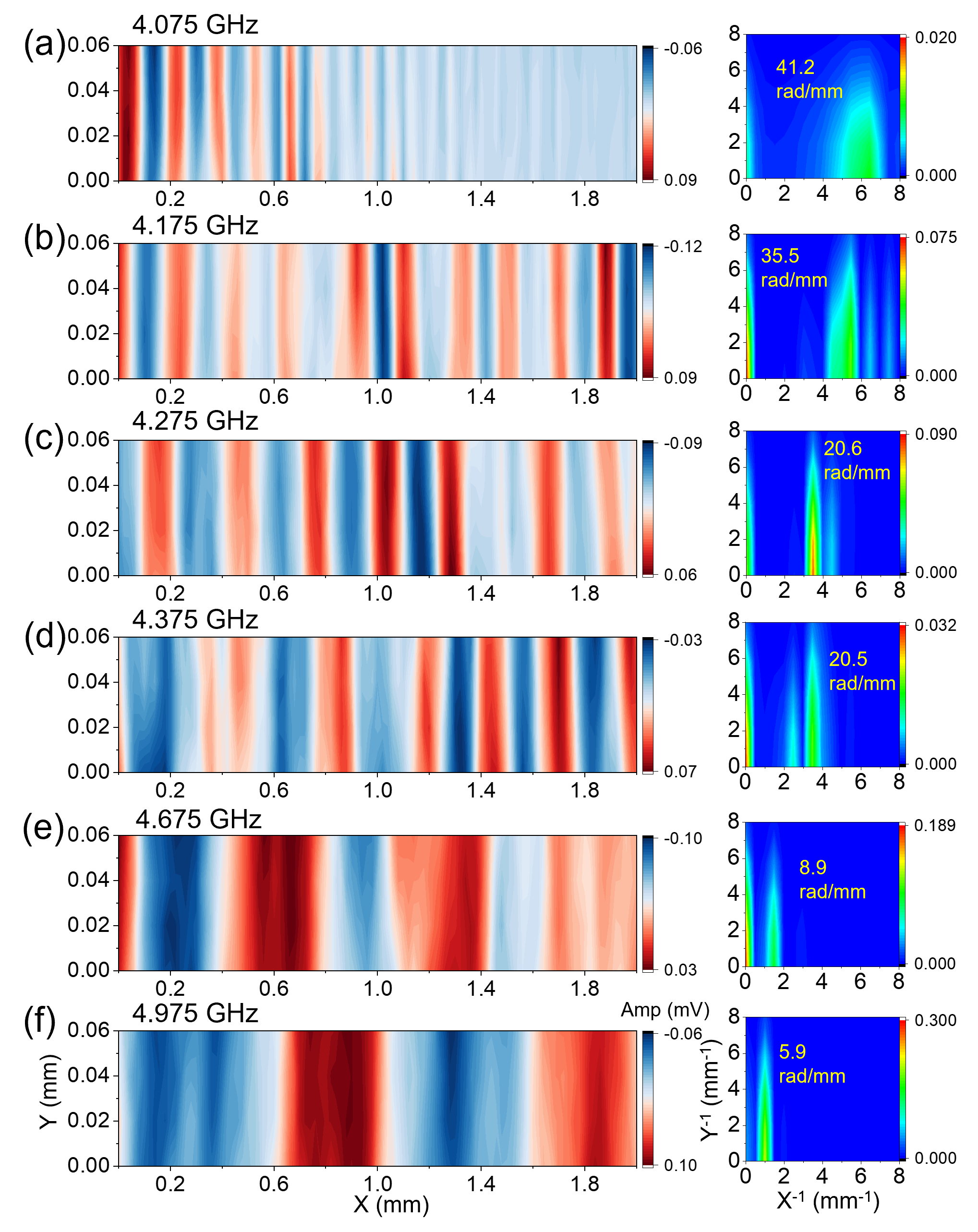}
 \caption{Scanned 2-D wavefront maps at fixed magnetic field $H=152.8$ mT (Left panel) and the corresponding 2-D FFT maps (right panel), at selective frequencies: (a) 4.075, (b) 4.175, (c) 4.275, (d) 4.375, (e) 4.675, and (f) 4.975. (unit in GHz). The extracted wavevectors are in the range between 5 -- 42 rad/mm.  }
 \label{fig:Fig4_2}
\end{figure}

For instance, Fig. \ref{fig:Fig4_2} exemplifies the 2-D wavefront maps measured at a fixed magnetic field $H=152.8$ mT and selective frequencies in the range of 4 - 5 GHz. Increasing the frequency from 4.075 GHz, Fig.\ref{fig:Fig4_2}(a) to 4.975 GHz, Fig.\ref{fig:Fig4_2}(f) \textcolor{black}{increases the spatial oscillation period, indicating an increasing wavelength and decreasing wavenumber, in agreement with the dispersion of BWVSWs based on the excitation configuration.} The corresponding wavevectors can be derived by performing 2-D FFT on the spatial maps, shown in the right panel of Fig.\ref{fig:Fig4_2}. The calculated central wavevectors are labeled next to each sub-figure in Fig.\ref{fig:Fig4_2}.

\textcolor{black}{In addition, we also noted that the spin wave amplitude exhibit variations across the scanned area. Since the excited BWVSWs propagate in the X axis, along the same direction of the CPW signal line, the CPW is radiating rf fields to the entire scanned area, instead of like a "point source". The spatial nonuniformity of the rf field may cause variations in its coupling to different modes, with maximum amplitudes occurring at different spatial locations. Such an effect is more pronounced for wavelengths that are much smaller than the CPW dimension, e.g. in Fig.\ref{fig:Fig4_2}(a-c).}

\subsubsection{Phase resolving -- FMR regime}

\begin{figure}[htb]
 \centering
 \includegraphics[width=3.5 in]{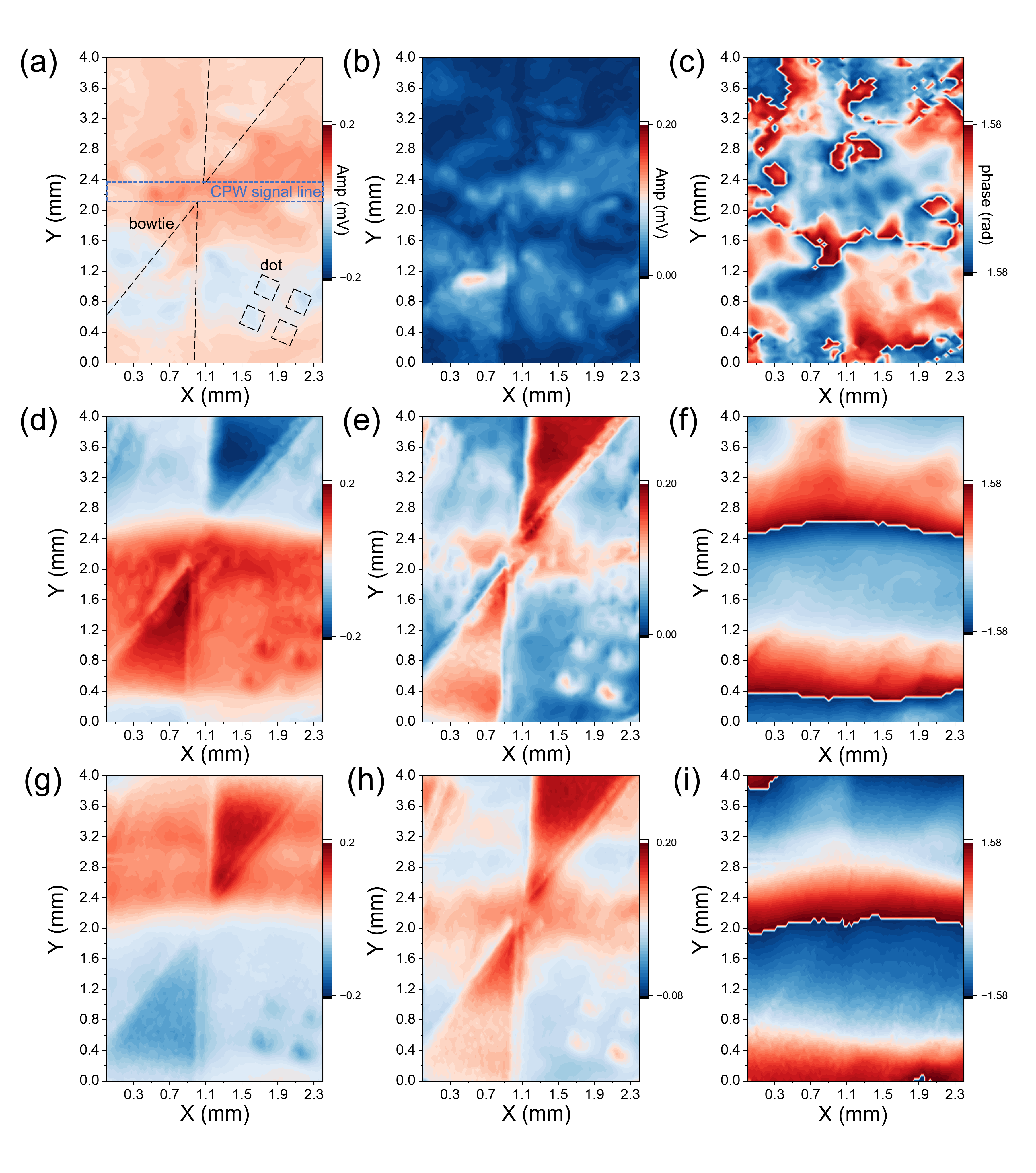}
 \caption{Scanned 2-D maps of the phase-resolving spin dynamics of a patterned Py/YIG bilayer by using the strobe light probe. A bowtie Py structure is centered around the CPW's signal line. Maps of the lock-in $X$ (a,d,g), amplitude (b,e,h), and phase (c,f,i) signals are shown for each magnetic field, $H$, and detection phase, $\phi_{rf}$. (a-c) At $H=0$ mT, and $\phi_{rf}=0^\circ$, off-resonance. (d-f) At $H=43.0$ mT, and $\phi_{rf}=0^\circ$, before the onset of the Py resonance. (g-i) At $H=43.0$ mT, but with swapped phase, $\phi_{rf}=180^\circ$. }
 \label{fig:Fig5}
\end{figure}

Next, we patterned the Py layer in the shape of a large bowtie structure centered on the YIG disc as well as adjacent microdot arrays using photolithography. We then deposited an additional Pt-layer (50-nm) covering the Py structure that serves as a mirror layer and enhances the overall light reflection. The patterned Py/YIG sample is again chip-flipped atop the CPW, with the signal line aligned approximately to the center of the bowtie, indicated in Fig.\ref{fig:Fig5}(a). We performed 2-D scans (along $x-y$) in the vicinity of the CPW.

\begin{figure}[htb]
 \centering
 \includegraphics[width=2.3 in]{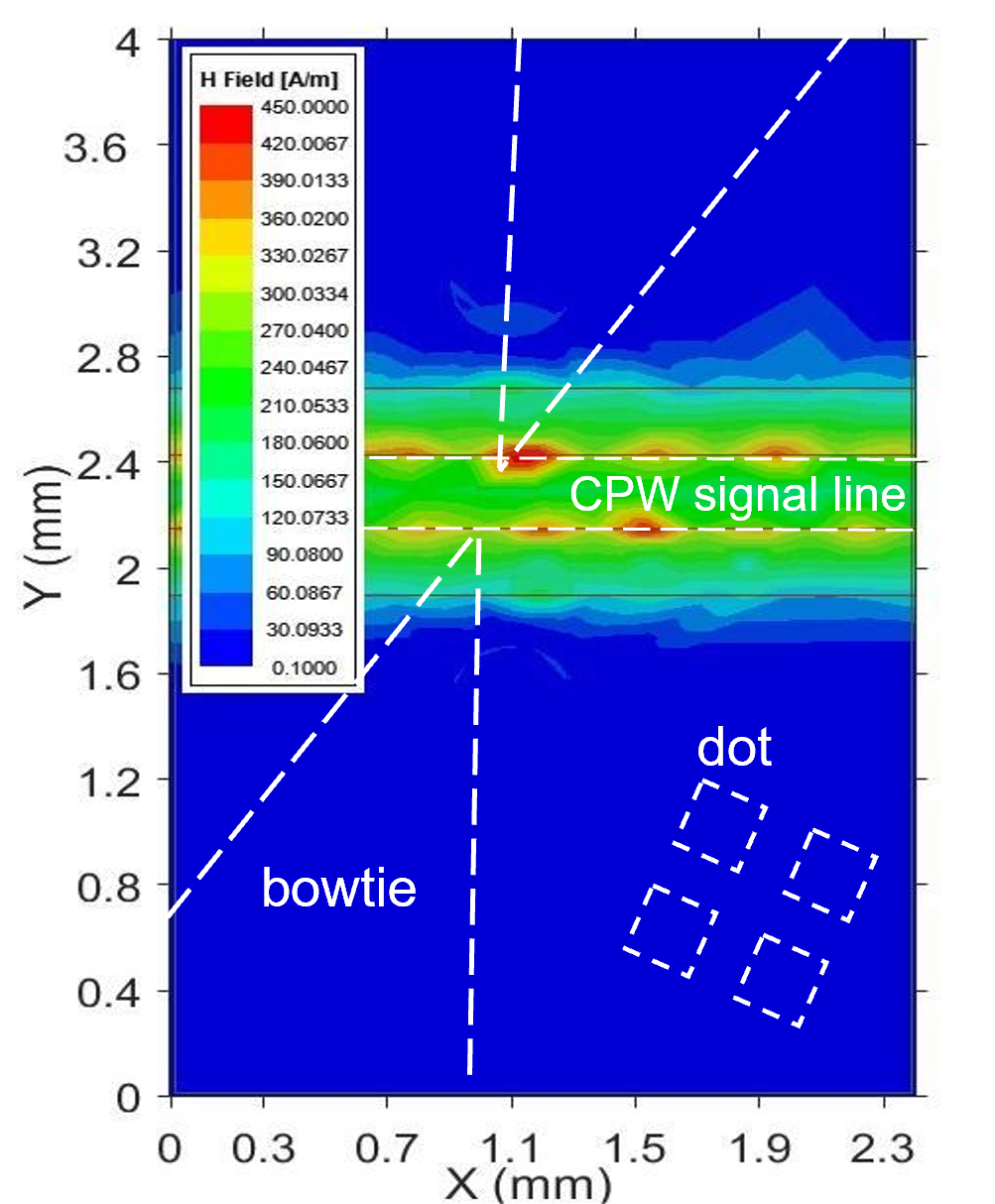}
 \caption{\textcolor{black}{Simulation and modeling of the microwave fields in the vicinity of the CPW structure. The CPW has a signal-line width of 0.28 mm, a gap of 0.25 mm. The dielectric layer is 0.356 mm, and the permittivity is 11.} }
 \label{fig:Fig5.5}
\end{figure}

\begin{figure*}[htb]
 \centering
 \includegraphics[width= 6.5 in]{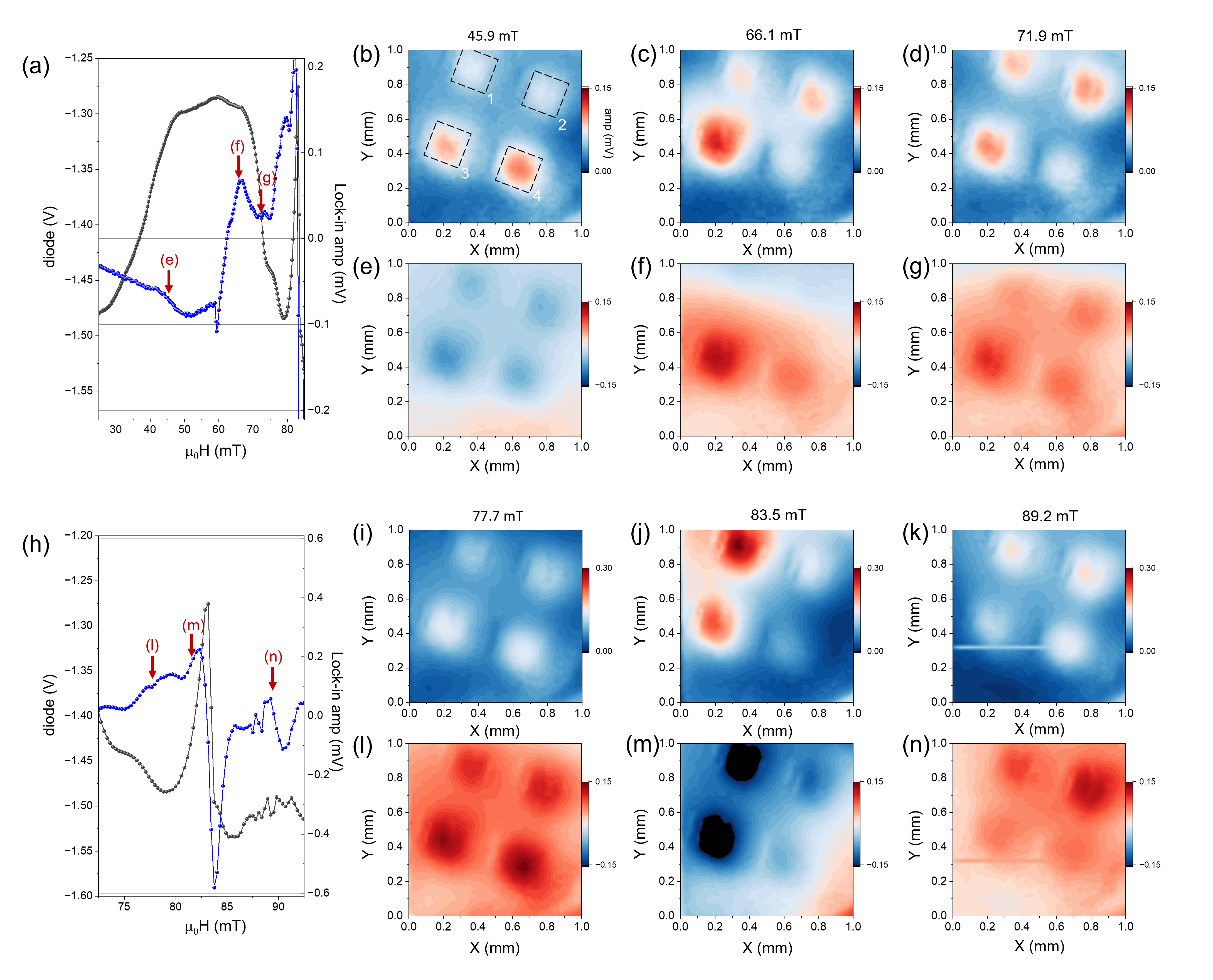}
 \caption{(a) Field-scan signal traces near the Py FMR at 4-GHz simultaneously probed by a rf power diode (left) and the lock-in $X$ (right). Three field points, at 45.9, 66.1, 71.9 mT (indicated by the arrows), were picked out for the spatial mapping of the spin dynamics in the vicinity of the $2 \times 2$ microdot array, labeled 1 -- 4. (b,c,d) Maps of the amplitude and (e,f,g) maps of the lock-in $X$ at the three respective fields. (h) Field-scan signal traces near the YIG FMR at 4-GHz simultaneously probed by a rf power diode (left) and the lock-in $X$ (right). Three field points, at 77.7, 83.5, 89.2 mT (indicated by the arrows), were picked out for the same spatial mapping. (i,j,k) Maps of the amplitude and (l,m,n) maps of the lock-in $X$ at the three respective fields. }
 \label{fig:Fig6}
\end{figure*}

\begin{figure*}[htb]
 \centering
 \includegraphics[width=7.1 in]{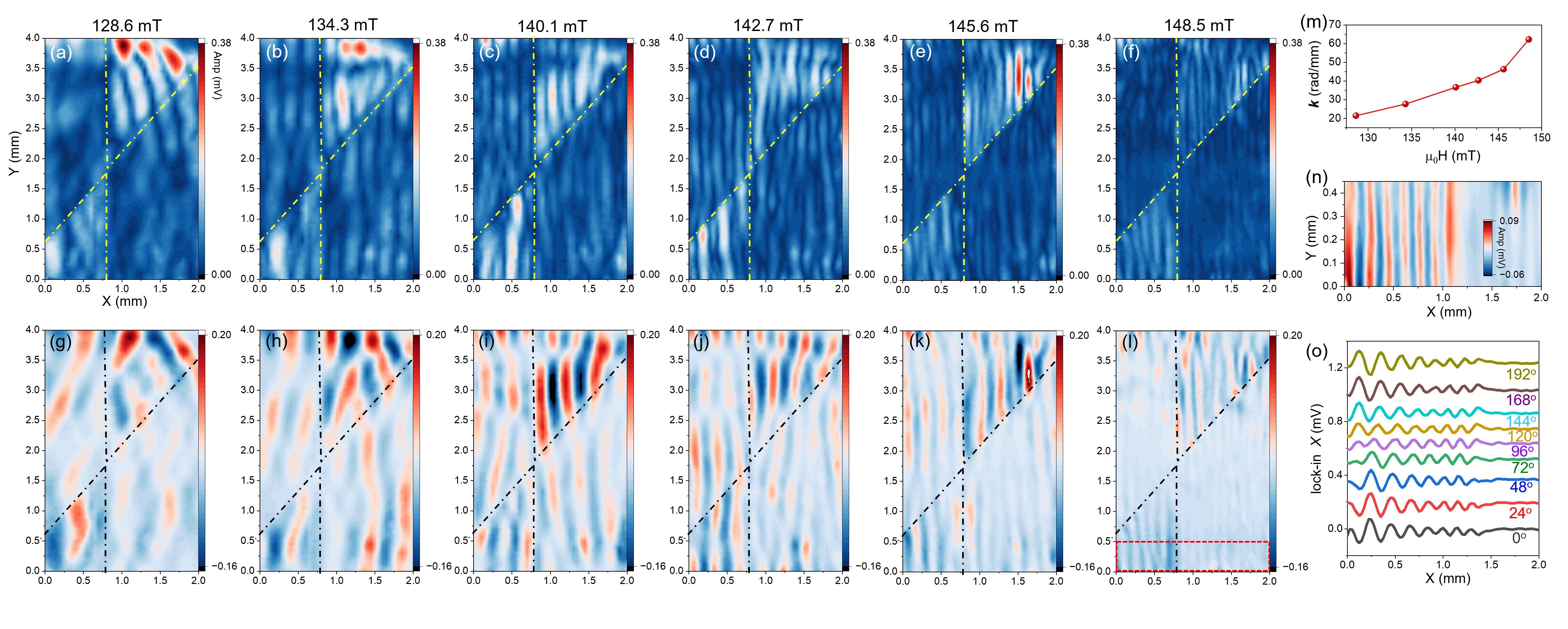}
 \caption{The scanned 2-D intensity (a-f) and wavefront (g-l) maps at selective magnetic fields, $H=$128.6, 134.3, 140.1, 142.7, 145.6, 148.5 mT. The yellow dashed line traces out the patterned Py bowtie structure. (m) The field-dependent wavevectors versus the frequency, extracted by FFT analysis of the 2-D wavefront maps. (n) A fine-scan window near the bottom bowtie at $H=148.5$ mT at a detection phase of 0$^\circ$. (o) 1-D scanned $x$-trace along $y=0$ mm and at different detection phases. The wavefront can be continuously tuned by the relative phase of the driving microwave. }
 \label{fig:Fig7}
\end{figure*}

By selecting the applied field, frequency, and detection phase (phase shifter), the images can be made sensitive to precession amplitude and phase. We show the scanned 2-D maps of the lock-in's $X$, amplitude, $\sim \sqrt{X^2+Y^2}$, and phase, arctan$(\frac{Y}{X}) = \phi_{eo}+\phi_{rf}-\phi_m$, at selective magnetic field and detection phase in Fig.\ref{fig:Fig5}. In our subsequent measurement, the frequency is set at 4-GHz and the detection phase $\phi_{rf}$ is set to a constant value during each scan, and the instrumental phase $\phi_{eo}$ varies only less than 0.2 rad across our 2-D scanned area. Therefore, the major contribution of the phase map comes from the magnetic phase $\phi_m$.  

At off-resonance ($H=0$ mT), the lock-in $X$ and the amplitude exhibit trivial signal, Fig.\ref{fig:Fig5}(a) and (b). The phase map, Fig.\ref{fig:Fig5}(c), exhibit random phase contrast. This indicates the decoupled magnetization and the driving microwave at the off-resonance condition. Thus the magneto-optical contrast is mainly from the random spin polarization states caused by unsaturated Py and YIG micro-structures. 

We then tune the magnetic field to $H=43.0$ mT (right before the onset of the Py resonance) and the detection phase $\phi=0^\circ$. Strong signals at the Py bowtie structure relative to the background offset were observed, indicating the excited spin dynamics. Notably, compared to the amplitude map in Fig. \ref{fig:Fig5}(e), the lock-in $X$ signal map in Fig.\ref{fig:Fig5}(d) exhibits a strong color contrast between the top (negative peak) and bottom (positive peak) sections, which is attributed to the out-of-phase spin dynamics for the top and bottom Py sections caused by the opposite rf driving field above and below the central signal line. Such a phase transition near the signal line is also manifested by the phase map in Fig.\ref{fig:Fig5}(f). Further, the quasi-uniform phase variation along the horizontal direction, compared to the off-resonance case in Fig.\ref{fig:Fig5}(c), indicates that the magnetization is saturated and coupled to the microwave driving field. Under the same magnetic field, when the detection phase $\phi_{rf}$ is changed to 180$^\circ$, the lock-in $X$ signal polarity is also reversed -- a positive/negative peak is found at the top/bottom Py section, respectively, relative to the background offset, Fig.\ref{fig:Fig5}(g), despite that the maps of amplitude, Fig.\ref{fig:Fig5}(h), and phase, Fig.\ref{fig:Fig5}(i), remain nearly unchanged.         

Next, we examine the magnetic-field-driven phase evolution in the vicinity of the Py and YIG FMR. To do this, we focus on the $2 \times 2$-microdot array that is away from the CPW's signal line so that the rf field is relatively uniform across the scan area. \textcolor{black}{To confirm the rf field distribution near the dot regime, we performed simulation and modeling of our CPW structure using the High Frequency Structure Simulator (HFSS) from ANSYS, see Fig. \ref{fig:Fig5.5}. The CPW has a signal-line width of 0.28 mm, a gap of 0.25 mm. The dielectric layer is 0.356 mm, and the permittivity is 11. Since the dot array is well below the signal line (into the bottom ground pad), the rf field is uniform across the whole regime.} The spin-wave imaging is then conducted at this regime, and the results are summarized in Fig.\ref{fig:Fig6}. The four dots are identified as 1 - 4 in Fig.\ref{fig:Fig6}(b). We set the frequency at 4-GHz and a detection phase $\phi_{rf}=180^\circ$ for the measurement. The field-scan FMR signal was probed concurrently using the strobe optics (lock-in $X$, measured in mV) and a rf diode (rectified voltage, in V). Figure \ref{fig:Fig6}(a) plots the field-scan traces near the Py FMR at 4-GHz. The rf diode signal measures a broad absorption profile, while the lock-in $X$ traces the precessional phase evolution. We picked three magnetic fields of interest, at 45.9, 66.1, 71.9 mT, and scanned the 2-D maps of lock-in $X$ and amplitude in the vicinity of the $2 \times 2$ microdot array.

First, the intensity maps directly reflect the spin precession amplitude,  Fig.\ref{fig:Fig6}(b-d). We observe that each dot peaks its resonance at different magnetic field values: at $H=45.9$ mT, dots 3 and 4 are first driven into strong resonance while dots 1 and 2 are less pronounced. As the field increases, the precession amplitude of dots 1 and 2 gradually emerge, while dot 4 becomes dim, Fig.\ref{fig:Fig6}(d). Nevertheless, all four dots are excited with decent precession amplitudes in this field range, due to the relatively large linewidth of Py -- also evidenced by the broad profile shown in Fig. \ref{fig:Fig6}(a). 

In contrast, the lock-in $X$ signal strongly differentiates the precession phase relative to the microwave drive. At $H=45.9$ mT, Fig.\ref{fig:Fig6}(e), all four dots are tuned to precess anti-phase with respect to the microwave drive (dark amplitude); at $H=71.9$ mT, Fig.\ref{fig:Fig6}(g), they are tuned to precess in phase with the microwave (bright amplitude); at the intermediate state ($H=66.1$ mT), Fig.\ref{fig:Fig6}(f), dots 3 and 4 are tuned in phase while dots 1 and 2 are tuned out of phase, therefore become `invisible' in the lock-in $X$ map, despite their strong emergence in the amplitude map.  

Next, we focus on YIG FMR regime, Figure \ref{fig:Fig6}(h) plots the field-scan traces near the YIG FMR at 4-GHz. The signal is notably stronger than the Py counterpart in Fig.\ref{fig:Fig6}(a), primarily due to the detection of the magneto-optical Faraday effect and the substantially thick YIG disc. We picked three magnetic fields in the vicinity, at 77.7, 83.5, 89.2 mT, and scanned the 2-D maps of the amplitude, Fig.\ref{fig:Fig6}(i,j,k), and the lock-in $X$, Fig.\ref{fig:Fig6}(l,m,n). Likewise, by selecting different bias fields, the images can be made sensitive to the precession amplitude and phase. At $H=77.7$ mT, all four dots start to emerge into resonance, see amplitude map in Fig.\ref{fig:Fig6}(i), and the phase map indicates their precession are in phase with the external microwave (bright contrast), Fig.\ref{fig:Fig6}(l). In addition, the amplitude of dots 3 and 4 is stronger than in 1 and 2. When the field is tuned to 83.5 mT, their precession are driven into an anti-phase state (dark contrast), see Fig.\ref{fig:Fig6}(m), and the amplitude of dots 1 and 3 is made significantly larger than dots 2 and 4 (that is nearly out of phase with the microwave), Fig.\ref{fig:Fig6}(j). The in-phase resonance can be recovered by further increasing the magnetic field to $H=89.2$ mT, Fig.\ref{fig:Fig6}(n). Contrary to Fig.\ref{fig:Fig6}(i), the amplitude map shows that dots 1 and 2 becomes stronger than in dots 3 and 4. Therefore, despite the nearly identical global microwave drive, the spin precessional state can vary largely in phase between isolated magnetic entities, and the strobe light detection can effectively and concurrently distinguish such phase variations in addition to their amplitude/intensity.  

\begin{figure*}[htb]
 \centering
 \includegraphics[width=7.1 in]{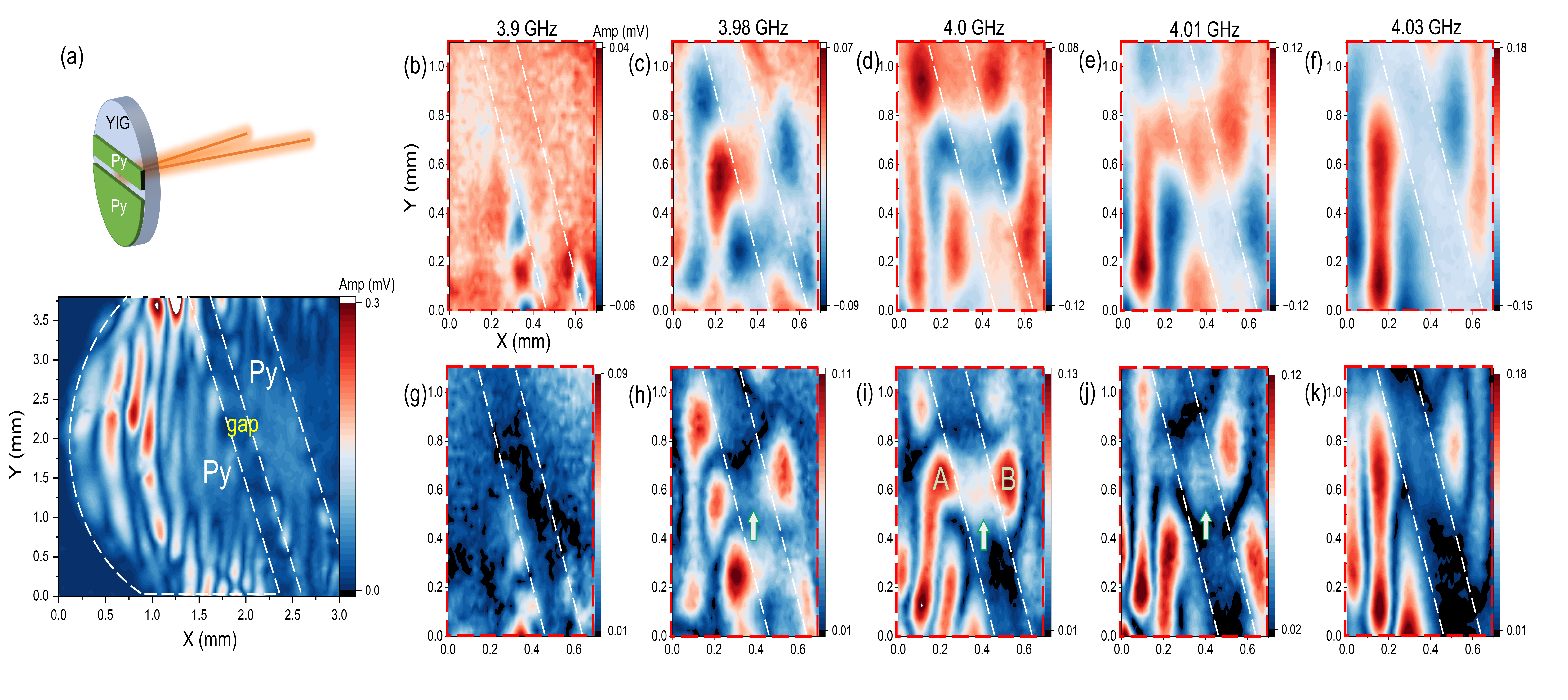}
 \caption{(a) The Py/YIG sample with a thin gap ($200$ $\mu$m-wide), and the 2-D scanned map (full-view) showing the spin wave excitation at 4 GHz and $H=149.4$ mT. The fine-scanned 2-D wavefront (b-f) and intensity (g-k) maps at selective frequencies, $f=3.9, 3.98, 4.0, 4.01, 4.03$ GHz, at $H=150.0$ mT, in the vicinity of the gap, showing pronounced spin wave hotspots near the edges of the Py underlayer. } 
 \label{fig:Fig8}
\end{figure*}

\subsubsection{Phase resolving -- YIG spin-wave regime}

Moving forward to discussions in the well-defined long wavelength regime ($H>120.0$ mT), we show the scanned 2-D intensity maps, in Fig.\ref{fig:Fig7}(a-f) and corresponding wavefront maps, in Fig.\ref{fig:Fig7}(g-l) at selective magnetic fields at 4-GHz. For the intensity maps, the signal strength is overall enhanced in areas atop the Py underlayer (bowtie structure) due to a stronger retroreflection from a thicker metal underlayer.

At $H=$128.6 -- 134.3 mT, the spin wave is still in transition between the caustic and the long wavelength regimes. Nontrivial $y$-component of the wavevector can be seen from the intensity map, Fig.\ref{fig:Fig7}(a-b). The wavefront bends towards the edge of the bowtie. The hot spots near the top edge can be attributed to the inhomogeneous internal field profile where a local reduction in the effective field causes a shift towards lower fields. The caustic behavior appears more significant in the wavefront map, Fig.\ref{fig:Fig7}(g-h), where the contrast of the bowtie underlayer blends into the caustic wave pattern.  

As the magnetic field increases, the caustic effect vanishes due to passing over the inflection point in the slowness curve, and the wavefronts become less prone to the underlayer-induced, effective demagnetization field. Therefore, the wavefront propagation becomes more and more dominant along the $x$-direction with reduced wavelengths compared to the size of the bowtie underlayer, see Fig.\ref{fig:Fig7}(c-f). The bowtie underlayer becomes also more apparent in the wavefront maps, Fig.\ref{fig:Fig7}(i-l). 

By performing the 2-D FFT analysis on the wavefront maps, we extract the field-dependent wavevector at the fixed frequency of 4-GHz, and the result is plotted in Fig.\ref{fig:Fig7}(m). In addition, similar to the earlier FMR case, the phase of the driving microwave can have a direct impact on the scanned wavefront map. Figure \ref{fig:Fig7}(n) shows a small fine-scan window, in the vicinity of the bottom bowtie (dashed enclosure in Fig.\ref{fig:Fig7}(l)) with a detection phase of 0$^\circ$. We then scanned 1-D $x$-traces at fixed $y$ = 0 mm with varying detection phase from 0$^\circ$ to 192$^\circ$, shown in Fig. \ref{fig:Fig7}(o). The wavefront can be continuously tuned from in-phase to anti-phase in a similar fashion demonstrated for the FMR scenario.       

\subsubsection{Collective dynamics near the underlayer edge}

The dipolar spin waves that are of interest in the present study are sensitive to additional magnetic entities producing local field variations in the vicinity, such as magnetic foreign objects, e.g. MFM tip/probe, magnetic nanoparticle, patterned magnetic structures, e.g. synthetic meta-surfaces, and magnetic underlayers \cite{kempinger2021field,montoncello2023brillouin,wang2023observation,negrello2022dynamic,dion2024ultrastrong}. For example, near the edge of the magnetic underlayer, such as in Fig. \ref{fig:Fig7}, spin wave `hotspots' can result due to edge excitations bestowing a large spin precession amplitude. Note that the Py saturation magnetization is about five times larger than that for YIG, so the Py underlayer can perturb the effective magnetic field in the YIG layer, despite the YIG itself being a continuous film. 

Such a response of the YIG local precession to the Py underlayer effectively enters the $\phi_m$ contribution in the strobe light detection. These high amplitude edge hotspots can also constructively or destructively interfere, governed by their precessional phase, and influence the spatial intensity distribution of the resultant magnon intensity, i.e., potentially controlling the energy and information flows associated with spin waves \cite{zhang2019controlled,lara2017information,caso2022edge}. 

To examine this effect, we covered half of the YIG disc with a 50-nm Py (in the shape of a semicircle), and ion-milled to form a thin gap in the shape of a strip. The gap is small enough as compared to the wavelength of the excited magnons. The scanned 2-D map in Fig. \ref{fig:Fig8}(a) shows the full-view amplitude map at 4 GHz and under a magnetic field $H=149.4$ mT (applied along the horizontal direction $X$). At this frequency and magnetic field, spin wave hotspots emerge at both edges of the Py underlayer in the vicinity of the thin gap. We focused on a small area near the gap and scanned the 2-D maps of wavefront, Fig.\ref{fig:Fig8}(b-f), and intensity, Fig.\ref{fig:Fig8}(g-k), at a fixed magnetic field, $H=150.0$ mT and selective frequencies. 

At $f=3.9$ GHz, the dispersion is beyond the cutoff regime thus no waves are excited. As $f$ increases to 3.98 GHz, the edge hotspots form on respective Py edges in the vicinity of the gap. However, the intensity remained weak inside the gap regime, Fig.\ref{fig:Fig8}(h). This is likely due to the anti-phase precession observed in the corresponding wavefront map, Fig.\ref{fig:Fig8}(c). At $f=4.0$ GHz, besides the two edge hotspots, `A' and `B', significantly enhanced excitation was also found inside the gap regime, channeling the two edge spots, Fig.\ref{fig:Fig8}(i). We attribute this to the in-phase precession near the two hotspots, `A' and `B', as indicated by the wavefront map, Fig.\ref{fig:Fig8}(d). Such an in-phase, collective resonance persists at $f=4.01$ GHz, Fig.\ref{fig:Fig8}(e), giving rise to gap excitation, Fig.\ref{fig:Fig8}(j). It once again vanishes, Fig.\ref{fig:Fig8}(k), when the edge hotspots become anti-phase, at 4.03 GHz, Fig.\ref{fig:Fig8}(f). Therefore, distinct mode excitation (in the gap) via coupled resonance (edge spots) can be annihilated or enhanced by destructive or constructive interference, governed by the phase relationship at the respective boundaries.  

\section{Summary}

In summary, we demonstrate phase-resolving spin wave microscopy using IR strobe light operating at 1550-nm. The method uses a cw fiber laser that is amplitude-modulated at the spin dynamic frequencies, and thus allows for coherent tracking of the spin precession phase. Using such a strobe light probe, the detected precessional phase contrast can be further employed to construct the spin wave wavefront \textcolor{black}{in the cw regime of spin wave propagation} without needing any optical reference paths. \textcolor{black}{We showcase spectroscopic studies and spatial mapping of the BWVSW modes in the dipolar wave regime, and in both continuous films and patterned samples. Key dispersion features, including the wavevector cutoff and spin wave caustics are observed, and the results are compared with that of using other probing techniques.}  

\textcolor{black}{The strobe feature of the optical probe allows to explicitly distinguish and quantify the various phase contributions, such as from the geometry \cite{xiong2020detecting}, spin-orbit torque \cite{li2019simultaneous}, and magnon-magnon interactions \cite{xiong2020probing,xiong2022tunable}. In particular, due to the phase accumulation arising from the spin wave propagation, the spin wave group velocity can be directly extracted from phase map of the dispersion. This capability opens up new opportunities in studying novel magnonic devices such as coupled magnon waveguides \cite{sadovnikov2022exceptional,an2024emergent}, magnonic cavities \cite{inman2022hybrid,santos2023magnon,liu2024strong}, and logic and network devices \cite{wang2024all,wang2024nanoscale,xiong2024magnon}}.    

The strobe light method shares the similar advantages with conventional lock-in based, field-sweep FMR measurements with high magnetic field resolution and broad dynamic range. Furthermore, the use of a cw laser and fiber-optics greatly simplifies the optical setup, allowing the system to be much less susceptible to external mechanical vibrations and noises, eliminating the possible optical artifacts associated with time-delay based techniques. \textcolor{black}{The relatively simple implementation also offers the potential of being made into a compact, tabletop system, integrating with common cryogenic and high-vacuum environments, and combining with other complementary fiber-based spectroscopy techniques \cite{li2023fiber,karimeddiny2023sagnac}}. 

The IR wavelength is suited for detecting both the magneto-optical Kerr and Faraday effects, thus enabling simultaneous probing of the spin dynamics of metallic (Py) and dielectric (YIG) bilayer systems. \textcolor{black}{Such a capability can find usefulness in studying hybrid magnonic systems, where phase tracking of coupled magnons (e.g. in magnetic multilayers) relative to other external excitations are of interest.} Beyond the investigation of magnon-photon and magnon-magnon coupling, such a wavelength was recently also found relevant in detecting phonon excitations \cite{taga2021optical,hisatomi2023quantitative,komiyama2024quantitative} in piezoelectric materials, thus opening up new prospects in studying coherent coupling between magnons and phonons, such as the bulk- and surface-acoustic wave driven magnonic phenomena \cite{li2021advances,liao2024nonreciprocal,shah2020giant,muller2024temperature,schlitz2022magnetization}.

\section{Acknowledgements}

The experimental work at UNC-CH was supported by Air-Force Office of Scientific Research (AFOSR) under award number FA2386-21-1-4091 and the U.S. National Science Foundation (NSF) under Grant No. ECCS-2246254. This work made use of instrumentation at the Chapel Hill Analytical and Nanofabrication Laboratory (CHANL), a member of the North Carolina Research Triangle Nanotechnology Network (RTNN), which is supported by the NSF, Grant ECCS-2025064, as part of the National Nanotechnology Coordinated Infrastructure, NNCI. D.S. acknowledges financial support from the National Science Foundation under Grant No. DMR-2143642. Y.L. and V.N. acknowledges support by the U.S. Department of Energy, Office of Science, Basic Energy Sciences, Materials Sciences and Engineering Division under Contract No. DE-SC0022060. T.H.K. acknowledges support by AFOSR under Grant No. NRF-2021K1A3A1A32084663.

\bibliography{sample}

\end{document}